\documentclass{ws-procs9x6}

\newcommand{\beq}{\begin{equation} }
\newcommand{\eeq} {\end{equation} }

\newcommand{\bed}{\begin{displaymath}}
\newcommand{\eed}{\end{displaymath}}

\begin{document}
\title{The low energy effective theory and  Nucleon Stability}

\author{Ilia Gogoladze}

\address{
Department of Physics
 University of Notre Dame,\\
  Notre Dame, IN 46556 USA}


\maketitle

\abstracts{ We show that the Standard Model Lagrangian, including
small neutrino masses, has an anomaly-free discrete $Z_6$
symmetry. Anomaly cancellation requires the  number of family to
be 3 $mod$ 6. This symmetry can ensure the stability of the
nucleon even when the threshold of new physics $\Lambda$ is low as
$10^2~ \mathrm{GeV}$. All $\Delta B=1$ and $\Delta B=2$ ($B$ is
the baryon number) effective operators are forbidden by the $Z_6$
symmetry. $\Delta B=3$ operators are allowed, but they arise only
at dimension 15. We suggest a simple mechanism for realizing
reasonable neutrino masses and mixings even with such a low scale
for $\Lambda$.}

The Standard Model (SM) has been highly successful in explaining
all experimental observations in the energy regime up to a few
hundred GeV. However, it is believed to be an effective field
theory valid only up to a cutoff scale $\Lambda$.
Non-renormalizable operators which are gauge invariant but
suppressed by appropriate inverse powers of $\Lambda$ should then
be considered in the low energy effective theory. The dimension 5
operator $\ell\ell H H/{\Lambda_L}$ ($\ell$ is the lepton doublet)
which violates lepton number ($L$) by two units is the lowest
dimensional of such operators. Experimental evidence for neutrino
masses suggests the effective scale of $L$-violation is around
$\Lambda_L\sim 10^{14}-10^{15}~\mathrm{GeV}$. The $d=6$ operator
$QQQ\ell/{ {\Lambda}^2_B}$ violates both baryon number ($B$) and
lepton number and leads to the decay of the nucleon. The current
limits on proton lifetime are $\tau_p
>5\times 10^{33}~\mathrm{yrs}$ for $p\rightarrow e^+ \pi^{0}$
\cite{pdg}. These limits imply that
$\Lambda_B>10^{15}~\mathrm{GeV}$. Grand Unified Theories with or
without supersymmetry generate such $B$-violating operator with
$\Lambda_B\sim 10^{15}-10^{16}~\mathrm{GeV}.$ These theories are
currently being tested through nucleon decay. Any new physics with
a threshold $\Lambda$ less than the GUT scale will thus be
constrained by both proton lifetime and neutrino masses.

As we know   the SM effective lagrangian does not have a
continuous anomaly-free symmetry that can suppress baryon number
and lepton number violating processes. This is our reason for
focusing on discrete symmetries. It is preferable that such
symmetries have a gauge origin\cite{kw} since all global
symmetries are expected to be violated by the quantum
gravitational effects. Discrete gauge symmetries have been
utilized in suppressing nucleon decay\cite{ibanez} as well as in
addressing other aspects of physics such as solving the $\mu$
problem\cite{axion} of supersymmetry, fermion mass hierarchy
problem\cite{wk} and the stability of the axion\cite{axion,Dias}.
A $Z_3$ baryon parity was found in Ref. [3]
 that suppresses nucleon decay. In order for it to
have a gauge origin, complicated particle content  were
introduced.

We pointed out the SM lagrangian has a discrete $Z_6$ gauge
symmetry which forbids all $\Delta B=1$ and $\Delta B=2$ baryon
violating effective operators. This can be seen as follows. The SM
Yukawa couplings incorporating the seesaw mechanism to generate
small neutrino masses is
\begin{eqnarray}
{ L}_{\rm Y} = Qu^{c}H+Qd^{c}{H}^*+\ell
e^{c}{H}^*+\ell\nu^{c}H+M_R \nu^{c}\nu^{c}~.
\end{eqnarray}
Here we have used the standard (lefthanded) notation for the
fermion fields and have not displayed the Yukawa couplings or the
generation indices. This lagrangian respects a $Z_6$ discrete
symmetry with the charge assignment as shown in Table 1.
\begin{table}[ht]
 \begin{center}
  {\renewcommand{\arraystretch}{1.1}
 \begin{tabular}{|c| c c c c c c c | }
  \hline
  \rule[5mm]{0mm}{0pt} & $Q$ & $u^c$ & $d^c$ & $\ell$ & $e^c$ & $\nu^{c}$& $H$ \\
  \hline
  \rule[5mm]{0mm}{0pt}
 $Z_6$&6 & 5 & 1 & 2 & 5 &3 & 1  \\
     \hline
\end{tabular}
 }\\
 \medskip
  {\footnotesize Table 1. Family-independent $Z_6$ charge assignment
   of the SM fermions and the Higgs boson. }
 \end{center}
\end{table}
From Table 1 it is easy to calculate the $Z_6$ crossed anomaly
coefficients with the SM gauge groups. We find the $SU(3)_C$ and
$SU(2)_L$ anomalies to be: $ A_{{[SU(3)_C]}^2\times Z_6}=3N_g$ and
$A_{{[SU(2)_L]}^2\times Z_6}=N_g$ where $N_g$ is the number of
generations. The condition for a $Z_N$ discrete group to be
anomaly-free is: $ A_i=\frac{N}{2}~~mod~N~$ where $i$ stands for
$SU(3)_C$ and $SU(2)_L$. For $Z_6$, this condition reduces to $
A_i=3~mod~6$, so when $N_g=3$, $Z_6$ is anomaly-free.  The
significance of this result is that unknown quantum gravitational
effects will respect this $Z_6$. It is this feature that we
utilize to stabilize the nucleon. Absence of anomalies also
suggests that the $Z_6$ may have a simple gauge origin.

We have found\cite{GBP} a simple and economic embedding of $Z_6$
into a $U(1)$ gauge symmetry associated with
$I^{3}_{R}+L_i+L_j-2L_k$. Here $L_i$ is the $i$th family lepton
number and $i\neq j\neq k$. No new particles are needed to cancel
gauge anomalies. With the inclusion of righthanded neutrinos
$I^{3}_{R}=Y-(B-L)/2$ is an anomaly-free symmetry. $L_i+L_j-2L_k$,
which corresponds to the $\lambda_8$ generator acting in the
leptonic $SU(3)$ family space, is also anomaly-free. The charges
of the SM particles under this $U(1)$ are:
 $Q_{i}=(0,0,0),~~~ {u_{i}}^c=(-1, -1, -1),~~~  {d_{i}}^c=(1, 1,
1),~~~
 l_{i}=(-4,2, 2),~~~  {e_{i}}^c=(5, -1, -1),~~~ {\nu_{i}}^c=(3, -3,
-3),~~~ H=1.$  This charge assignment allows all quark masses and
mixings as well as charged lepton masses. When the $U(1)$ symmetry
breaks spontaneously down to $Z_6$ by the vacuum expectation value
of a SM singlet scalar field $\phi$ with a charge of 6, realistic
neutrino masses and mixings are also induced\cite{GBP}.

From Table 1 it is easy to see that the $Z_6$ discrete  symmetry
allowed only $\Delta B=3$ effective operators with
lowest-dimension $d=15$ and forbids  all $\Delta B=1$  and $\Delta
B=2$  operators. $\Delta B=3$ and $d=15$  operator will lead to
``triple nucleon decay" processes where three nucleons in a heavy
nucleus undergo collective decays. We choose a specific operator
$Q^5\bar{d^c}^4\bar{\ell}/\Lambda^{11}$ as an example to study the
process $pnn\rightarrow e^{+}+\pi^{0}$ triple nucleon decay
process. In this case the triple nucleon decay lifetime can then
be estimated to be \beq \tau\sim
\frac{16\pi{f_{\pi}}^2{\Lambda}^{22}R^{6}}{ P^{2}\beta^{6}
M_{^{3}H}~},\eeq where $\beta\simeq 0.014~\mathrm{GeV}^{3}$ is the
matrix element to convert three quarks into a nucleon
\cite{jlqcd}, $f_{\pi}=139$ MeV is the pion decay constant, $P$ is
the probability for three nucleons in Oxygen nucleus to overlap in
a range the size of Tritium nucleus, R is the ratio between the
radii of Tritium nucleus and Oxygen nucleus. By putting the
current limit on proton lifetime of $3\times
10^{33}~\mathrm{yrs}$, we obtain: $ \Lambda\sim
10^2~\mathrm{GeV}.$ Thus we see the $Z_6$ symmetry ensures the
stability of the nucleon.

If the threshold of new physics is low as a few TeV, neutrino mass
induced through the effective operator $\ell\ell H H/\Lambda$
will be too large. We found a mechanism by which such operators
can be suppressed by making use of a discrete $Z_N$ symmetry (with
$N$ odd) surviving to low scale.

Consider the following effective operators in the low energy
lagrangian: \beq\label{cc} { L}\supset\ell\ell H
H\frac{S^{6}}{\Lambda^{7}}+\frac{S^{2N}}{{ \Lambda^{2N-4}}}~.\eeq
Here $S$ is a singlet field which has charge $(1,~3)$ under
$Z_N\times Z_6$ while $\ell$ has charge $(-3,~2)$. (The $Z_6$
charges of SM particles are as listed in Table~1.) In this case,
if $\Lambda=10~\mathrm{TeV}$ and $S=10^2~\mathrm{GeV}$, the
neutrino mass is of order $O(0.1)~\mathrm{eV}$, which is
consistent with the mass scale suggested by the atmospheric
neutrino oscillation data.

Two explicit examples of the $Z_N$ symmetry with $N=5$ and $7$ are
shown in Table 2. These $Z_N$ symmetries are free from gauge
anomalies. In the $Z_5$ example, the crossed anomaly coefficients
for $SU(3)_C$ and $SU(2)_L$  are $5N_g$ and $5N_g/2$ respectively
showing that $Z_5$ is indeed anomaly-free. For $Z_7$, these
coefficients are $7N_g$ and $7N_g/2$, so it is also anomaly-free.
\begin{table}[ht]\label{nu}
 \begin{center}
  {\renewcommand{\arraystretch}{1.1}
  \begin{tabular}{|c| c  c  c  c c   cc |   }
     \hline
     \rule[5mm]{0mm}{0pt}
   Field & $Q$ & $u^{c}$ &
   $d^{c}$ & $\ell$ &
   $e^{c}$  &
   $H$  & $S$\\
   \hline
   $Z_5$ & 1 & 4 & 4 &
   2 & 3  & 0  & 1\\
   $Z_7$ & $1$ & 6 & 6 &
   4 & 3  & 0  & 1\\
   \hline
  \end{tabular}
  }\\
  \medskip
 {\footnotesize  Table 2.~~$Z_N$ charge assignment for $N=5$ and $7$.}
  \label{z7}
 \end{center}
\end{table}

It is interesting to ask if the $Z_N$ can be embedded into a
gauged $U(1)$ symmetry. A simple possibility we have found is  to
embed this $Z_N$ into the anomalous $U(1)_A$ symmetry of string
origin with the anomalies cancelled by the Green-Schwarz mechanism
\cite{gs}. Consider $U(1)_{B-L}$ without the right handed
neutrinos but with the inclusion of vector-like fermions which
have the quantum numbers of ${\bf 5}(3)$ and ${\bar{\bf 5}}(2)$
under $SU(5)\times U(1)_A$. This $U(1)_A$ is anomaly-free by
virtue of the Green-Schwarz mechanism. When this $U(1)_A$ breaks
down to $Z_5$, the extra particles get heavy mass and are removed
from the low energy theory which is the $Z_6\times Z_5$ model.

This work  was supported in part by the National Science
Foundation under grant PHY00-98791.


\end{document}